\documentclass[useAMS,usenatbib]{mn2e}
\usepackage{graphicx}
\usepackage{epsfig}
\usepackage{amssymb}
\usepackage{amsmath}
\usepackage{psfrag}

\newcommand{\B}[1]{\mathbf{#1}}
\newcommand{\Br}{\B{r}}

\newcommand{\Bx}{\B{x}} 
 
\newcommand{\Bv}{\B{v}}

\newcommand{\Eq}[1]{Eq.~(\ref{#1})} 
\newcommand{\Fig}[1]{Fig.~\ref{#1}}
\newcommand{\Sec}[1]{Sec.~\ref{#1}}
\newcommand{\EqDef}{\equiv}

\newcommand{\tw}{\tilde{\omega}}

\title{Inconsistency in theories of violent-relaxation}
\author[I.~Arad and D.~Lynden-Bell]{I.~Arad\thanks{Email: 
  arad@ast.cam.ac.uk} and D.~Lynden-Bell\\
  Institute of Astronomy, Madingley Road, Cambridge CB3 OHA, UK}
\begin{document}

\date{Version of \today}
\pagerange{\pageref{firstpage}--\pageref{lastpage}} \pubyear{2004}

\maketitle

\label{firstpage}

\begin{abstract}
  We examine an inconsistency in theories of violent-relaxation by
  Lynden-Bell and Nakamura. The inconsistency arises from the
  non-transitive nature of these theories: a system that under goes a
  violent-relaxation, relaxes and then upon an addition of energy,
  under goes violent-relaxation once again would settle in an
  equilibrium state that is different from the one that is predicted
  had the system would go directly from the initial to the final
  states. We conclude that a proper description of the
  violent-relaxation process cannot be achieved by equilibrium
  statistical mechanics approach, but instead a dynamical theory for
  the coarse-grained phase-space density is needed.
\end{abstract}


\section{Introduction}

The main purpose of this paper is to demonstrate that theories
predicting a definite statistical equilibrium, dependent only on the
energy and the initial volume $\tau(\eta)d\eta$ at phase-space
densities in the range $\eta\to \eta+d\eta$, do not predict the same
final state when the system undergoes two violent relaxation sessions
separated in time, as they do when the two sessions are treated as
one. This may be called an inconsistency or at best a lack of
transitivity.

Consider an $N$-body gravitating system that starts in some state far
from equilibrium, vibrates violently and settles to a dynamically
steady state $s_1$ that lasts long enough that it may be considered
the final product of violent relaxation with energy $E_1$. Now suppose
that this system suffers a significant tidal disturbance from a
passing object that causes violent vibrations and leaves out the
system to relax again but now with energy $E_2$. There are now two
ways of predicting the outcome. Either we take the function
$\tau_0(\eta)$ giving the volume at each phase-space density from the
initial state $s_0$, or we use $\tau_1(\eta)$ the predicted outcome
for that function after the first relaxation process. Of course if we
used the fine grained phase-space density both would be the same, but
by hypothesis the system lasted so long in state $s_1$ that is at
``equilibrium'', and only the coarse-grained density can any longer be
relevant to the dynamics of the final relaxation. We show that the
outcomes predicted with energy $E_2$ and volume functions
$\tau_0(\eta)$ and $\tau_1(\eta)$ are certainly different in the
Lynden-Bell theory of violent relaxation \citep{ref:Lyn67}, as well as
in a more recent theory by Nakamura \citep{ref:Nak00}.

This paper is organised as follows: in \Sec{sec:overview} we give a
short overview of some of the difficulties in the theory of violent
relaxation, which we feel complement the main subject of this paper.
Then in \Sec{sec:LB67} we demonstrate the non-transitivity of the
Lynden-Bell theory of violent relaxation, using the thought experiment
that was presented above. In \Sec{sec:info} we give a brief
description of Nakamura's theory which is based on the
information-theory approach.  We re-derive his theory using a
combinatorial approach that enables us to compare it to Lynden-Bell's
theory. Then in \Sec{sec:NK00} we demonstrate that also Nakamura's
theory is non-transitive, using the thought experiment once again. In
\Sec{sec:conc} we present our conclusions.

\section{An overview of the difficulties in the theory of violent relaxation}
\label{sec:overview}

In this section we offer a short discussion of other
problems connected with the process of violent-relaxation and
`theories' that aim to predict its outcome. Let us start from first
principles. Most large $N$-body systems governed by long-range forces
which are not initially in balance will oscillate with decreasing
amplitude before they settle into a state in which the potential of
the long range force becomes almost steady. Such violent-relaxation
processes are known to occur in gravitational $N$-body systems.
Thereafter evolution may continue due to the shorter range interaction
in which the graininess of the individual particles is of importance,
but there is a large class of systems in which this secondary
evolution is on a much longer timescale. Violent relaxation under
gravity does not last long. After a few oscillations on the timescale
$(G\bar{\rho})^{-1/2}$ it is over. Thus the whole idea that the
interaction of the particles with the mean field will lead to some unique
detailed statistical equilibrium state, only dependent on the initial
conditions via the dynamically conserved quantities, is more of a vain
hope and a confession of ignorance than an established fact.
Nevertheless, \citet{ref:Hen68} gave some evidence in favour of its
prediction and for cosmological initial conditions \citet{ref:Nav95,
  ref:Nav96, ref:Nav97} show a considerable universality in their
results. \citet{ref:Bin04} gives a lovely toy model which he uses to
criticise deductions from $N$-body simulations. Even in the initial
discussion \citep{ref:Lyn67} it was admitted that there were many
stable steady states into which a gravitating system could settle and
that violent relaxation would be incomplete so that the system would
not inevitably attain a state close to the more probable one.

A second worrying aspect of any equilibrium theory is the apparent
lack of an analogue to the law of detailed balance. At real
thermodynamic equilibrium there are no cyclic processes going around
and around, but each individual emission process is exactly balanced
by the corresponding absorption process. In radiation theory Einstein
introduced his stimulated emission process just to ensure that this
would be so. In the process of violent relaxation each element is
interacting with the potential of the whole system, and one might
expect some to be highly accelerated as Fermi argued to get his cosmic
ray acceleration process. Conservation of energy must however lead to
some dynamical friction term but this cannot be mass related as that
is at odds with our earlier arguments that energy gain is independent
of mass.

Finally, not all systems have violent relaxation. In some early
experiments with pulsating concentric spherical shells,
\citet{ref:Hen68} found a few examples of gravitating system with
persistent oscillation that defied the general decay. Newton in
Principia showed that systems with a force law between particles
proportional to separation (rather than inverse square) oscillated
forever. Indeed Newton solved that $N$-body problem completely,
showing that each particle moved on a central ellipse centred on the
barycentre and that all those orbits had the same period. Lynden-Bell
\& Lynden-Bell (1999a) showed that Newton's work could be extended to
all systems in which the total potential energy was any function
$V(r)$ where $r^2 = \sum_i m_i(\Br_i - \bar{\Br})^2/M=
\frac{1}{2}\sum_{i,j}m_i m_j(\Br_i-\Br_j)^2/M$ and $\bar{\Br}$ is the
position of the barycentre. Likewise, the $N$-particle Schrodinger
equation was exactly solved for such potentials \citep{ref:Lyn99b};
the most interesting example of which is probably $V=GM^2/r$. This has
a force between any two particles of $Gm_i m_j(\Br_i-\Br_j)/r^3$ with
the final $r^3$ not $|\Br_i - \Br_j|^3$ but the $r$ defined above. A
related exception to violent relaxation with a most interesting
ever-pulsating detailed ``Maxwellian'' statistical equilibrium is the
system with an inter-particle potential $\frac{1}{2}m_i m_j (kr_{ij}^2
+ k'r^{-2}_{ij})$. In spite of the complication of the individual
particle motions due to the inverse cubic repulsion between particles,
such systems pulsate in scale forever in exact undamped simple
harmonic motion \citep{ref:Lyn04}.

\section{Non-transitivity in the Lynden-Bell theory of violent
  relaxation}
\label{sec:LB67}

The Lynden-Bell theory of violent relaxation (hereafter LB67) aims at
predicting the final equilibrium state of a collisionless gravitating
system undergoing a violent relaxation. In order to demonstrate its
non-transitive nature we shall first recall its main results for the
general case where initially the system has more than one phase-space
density levels.

\subsection{Main results of the LB67 theory}
\label{sec:LB-res}

In the LB67 theory we deal with a system which is initially out of
equilibrium. Its initial state is specified by its total energy $E$
and the phase-space volumes $V_1, V_2, \ldots$ of the initial
phase-space density levels $\eta_1, \eta_2, \ldots$\footnote{In the
  original formulation of LB67, Lynden-Bell used the masses of the
  different levels instead of their phase-space volume. However these
  are trivially related to each other by $M_J = \eta_J V_J$}.
Phase-space is then divided into \emph{micro-cells} of fixed volume
$\tw$, which can be either empty or hold a phase-space element of one
of the prescribed levels. The state of all these micro-cells defines a
\emph{micro-state}.

Next, we let the system re-distribute its phase-space elements as it
approaches an equilibrium. The micro-cells are then grouped into
macro-cells, each macro-cell containing $\nu$ micro-cells. A
\emph{macro-state} of the system is defined by the matrix $\{n_{iJ}\}$
which specifies how many phase-space elements of type $J$ ended up in
the macro-cell $i$. The coarse-grained phase-space density function
(DF) at the macro-cell $i$ is therefore given by
\begin{equation}
  f_i = \frac{1}{\nu}\sum_J \eta_J n_{iJ} \ .
\end{equation}

Then, in the spirit of ordinary statistical mechanics, one assumes
that the system has an equal a priori probability of being in each one
of the micro-states. To find the equilibrium state one maximises the
function $W\big(\{n_{iJ}\}\big)$ which counts the number of
micro-states that correspond to the macro-state $\{n_{iJ}\}$, hence
obtaining the most probable state.

When maximising $W$ one has to consider only those macro-states for
which the total energy is $E$ and the overall volume in each of the
initial phase-space levels is $V_1, V_2, \ldots$. This can be done in
the usual way with Lagrange multipliers. After passing to a continuous
description of the macro-cells, the resultant DF is
\begin{equation}
\label{def:LB-f}
  f_{LB}(\Br,\Bv) = f_{LB}[\epsilon(\Br,\Bv)] 
     \EqDef A(\epsilon) \sum_J \eta_J 
      e^{-\beta\eta_J(\epsilon - \mu_J)} \ ,
\end{equation}
with 
\begin{equation}
\label{def:LB-A}
    A(\epsilon) \EqDef \frac{1}
       {1 + \sum_J e^{-\beta\eta_J(\epsilon - \mu_J)}} \ .
\end{equation}
Here $\epsilon(\Br,\Bv) = v^2/2 + \Phi_{LB}(r)$ is the energy per unit
mass of the $(\Br,\Bv)$ phase-space cell, $\Phi_{LB}(r)$ is the
gravitational potential, calculated self-consistently from the Poisson
equation
\begin{equation}
  \nabla^2 \Phi(r) = 4\pi G\int\!\! d^3\Bv\, f_{LB}(\Br,\Bv) \ .
\end{equation}
The dimensional constants $\beta, \mu_1, \mu_2, \ldots$ are the
Lagrange multipliers, which are calculated from the energy
conservation constraint
\begin{equation}
  \label{eq:LB-energy}
  \int\!\!d^6\tau\, f_{LB}(\Br,\Bv) 
     \left[\frac{v^2}{2} + \frac{1}{2}\Phi_{LB}(r)\right] = E \ ,
\end{equation}
and the initial conditions constraints
\begin{equation}
  \label{eq:LB-vol}
  \int\!\! d^6\tau\, A(\epsilon) 
     e^{-\beta\eta_J(\epsilon - \mu_J)} = V_J \ , 
      \quad J=1, 2, \ldots \ ,
\end{equation}
where we have used used the notation $d^6\tau \EqDef d^3\Br d^3\Bv$.

Equations~(\ref{def:LB-f}-\ref{eq:LB-vol}) are the main results of the
LB67 theory. For our needs, however, two small modifications are
needed. Firstly, by introducing the energy density function
\begin{equation}
\label{def:g}
  g(\epsilon_0) \EqDef \int\!\! d^6\tau \,
    \delta[\epsilon(\Br,\Bv)-\epsilon_0] \ ,
\end{equation}
the integral in the LHS of \Eq{eq:LB-vol} can be written as $\int\!\!
g(\epsilon) A(\epsilon) e^{-\beta\eta_J(\epsilon - \mu_J)}\,
d\epsilon$. Secondly, we pass to a continuous description of the
initial density levels using the phase-space volume distribution
function $\tau(\eta)$, which is defined so that $\tau(\eta)d\eta$ is
the phase-space volume initially occupied by phase-space densities in
the range $\eta\to \eta+d\eta$. Formally, if $f_I(\Br,\Bv)$ is the
initial DF then $\tau(\eta)$ is given by
\begin{equation}
  \tau(\eta) = \int\!\! d^6\tau\, \delta[f_I(\Br,\Bv) - \eta] \ .
\end{equation}
By letting each density level $\eta_J$ have a small width
$\Delta\eta$, the $\sum_J$ sums in Eqs.~(\ref{def:LB-f},
\ref{def:LB-A}) can be changed to integrals by
\begin{equation}
  \sum_J \to \frac{1}{\Delta \eta}\int\!\! d\eta \ ,
\end{equation}
hence Eqs.~(\ref{def:LB-f}-\ref{eq:LB-vol}) are now
\begin{eqnarray}
\label{eq:LB2-f}
  f_{LB}(\epsilon) &=& \frac{1}{\Delta\eta}A(\epsilon)
    \int\!\!\eta e^{-\beta \eta(\epsilon-\mu_\eta)}\, d\eta \ , \\
\label{eq:LB2-A}
  A(\epsilon) &=& \frac{1}
     {1 + \frac{1}{\Delta\eta}\int\!\! 
        e^{-\beta \eta(\epsilon-\mu_\eta)}\, d\eta} \ , \\
\label{eq:LB2-possion}
  \nabla^2 \Phi(r) &=& 4\pi G\int\!\! d^3\Bv\, f_{LB}(\Br,\Bv) \ , \\
\label{eq:LB2-E}
  E &=& \int\!\!d^6\tau\, f_{LB}(\Br,\Bv) 
     \left[\frac{v^2}{2} + \frac{1}{2}\Phi(r)\right] \ , \\
\label{eq:LB2-vol}
  \tau(\eta)\Delta\eta &=& \int\!\! g(\epsilon) A(\epsilon) 
      e^{-\beta \eta(\epsilon-\mu_\eta)}\, d\epsilon \ .
\end{eqnarray}
Notice that the amplitude of $\Delta\eta$ is unimportant as it can
always be absorbed into the Lagrange multipliers $\mu_\eta$.

Finally we recall that in a spatially infinite domain
equations~(\ref{eq:LB2-f}-\ref{eq:LB2-vol}) have no solution since the
density can spread indefinitely while conserving its energy and
increasing its entropy. A common way to overcome this problem, which
shall be adopted here, is to work within an rigid sphere of radius
$R$, and to assume that the resulting equilibrium configuration is
spherical. Such model, although not realistic, is an easy way to
obtain a finite solution.

\subsection{The double relaxation experiment}
\label{sec:double-relax}

To test the transitivity of the LB67 theory we propose the following
four-steps thought experiment which was mentioned in the
introduction:
\begin{enumerate}
\item We prepare a system with one density-level (the water-bag
  configuration) and a total energy of $E$. In accordance with the
  above section, we put the system inside a sphere of radius $R$.  We
  denote the initial state of the system by $s_0$.
\item We let the system go trough a violent relaxation process to an
  equilibrium which is denoted by $s_1$.
\item We add an amount of $\Delta E$ energy to the system by, for
  example, a strong impulse of an external gravitational field. The
  system then goes once again through a violent relaxation process and
  settles down in a new equilibrium state, denoted by $s_2$.
\item We prepare a new system with the same parameters as $s_0$ except
  for the energy, which is set to $E+\Delta E$. We let it go through a
  violent relaxation process to an equilibrium which is denoted by
  $s_3$.
\end{enumerate}
Now if the theory is transitive then necessarily $s_2 = s_3$. 

To see if this is really the case in the LB67 theory, we begin by
calculating the $s_1$ and $s_3$ states which are relatively simple to
calculate, being the outcome a water-bag configuration. This
calculation has been fully done in \citet{ref:Cha98a}, and here we use
some of their results.

The $s_0$ step was prepared with total mass $M=1, G=1, R=1$ and an
initial phase-space density level 
\begin{equation}
\label{eq:eta}
  \eta_0 = \frac{10^3}{\sqrt{512\pi^4 G^3 M R^3}} \ ,
\end{equation}
which according to \citet{ref:Cha98a} guarantees that for each energy
there would be only one equilibrium state.

From \Eq{def:LB-f} we find that the DF of $s_1$ and $s_3$ is the
well-known Fermi-Dirac distribution
\begin{eqnarray}
\label{eq:FD1}
  f_1(\epsilon) &=& \frac{\eta_0}
       {1 + e^{\beta_1\eta_0(\epsilon-\mu_1)}} \ , \\
\label{eq:FD2}
  f_3(\epsilon) &=& \frac{\eta_0}
       {1 + e^{\beta_3\eta_0(\epsilon-\mu_3)}} \ ,
\end{eqnarray}
with $\beta_1, \beta_3, \mu_1, \mu_3$ Lagrange multipliers to be fixed
from the energy constraint and the initial conditions. As there is
only one density-level, the initial condition constraint can be
replaced with the conservation of mass constraint. 

In \citet{ref:Cha98a} it is shown how the Lagrange multipliers can be
found for a given mass and energy, and we therefore do not repeat
these steps here but instead give the values of these parameters in
Table~1. All numerical calculations were done using the \emph{GNU
  Scientific Library 1.5} \texttt(GSL 1.5), which is a free software
available from \texttt{http://www.gnu.org/software/gsl/}. The
differential equation for $\Phi(r)$ was solved using an embedded
Runge-Kutta Prince-Dormand (8,9) method, whereas integration was done
using a 51 points Gauss-Kronrod rule. In all calculations a relative
error of less than $10^{-3}$ was maintained.

\begin{figure}
  \centerline{ \hbox{
      \epsfig{file=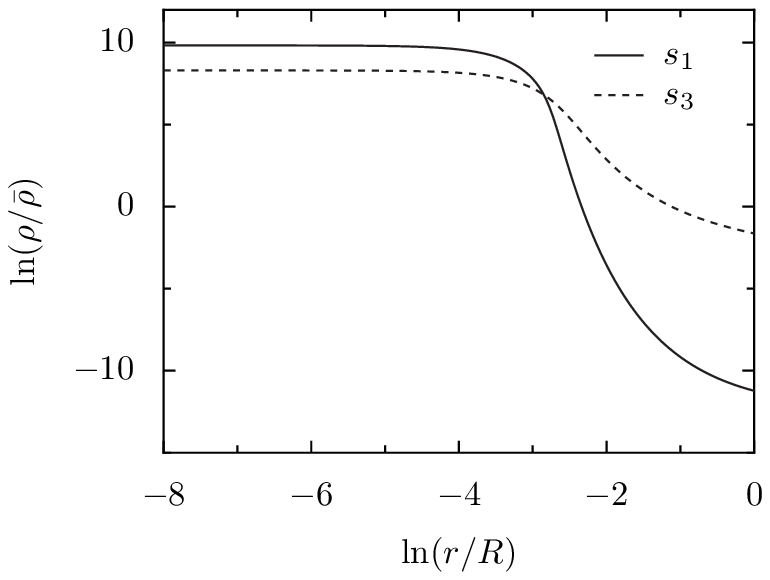,width=0.40\textwidth} } }
  \caption{Density profiles of the $s_1$ and $s_3$ states in the
    double-relaxation experiment of the LB67 theory. $\bar{\rho}$ is
    the average density given by $\bar{\rho}=3M/(4\pi R^3)$.  The
    $s_3$ state corresponds to a hotter system with higher energy
    which makes it less concentrated.}
  \label{fig:rhos-LB}
\end{figure}

\begin{figure}
  \centerline{ \hbox{
      \epsfig{file=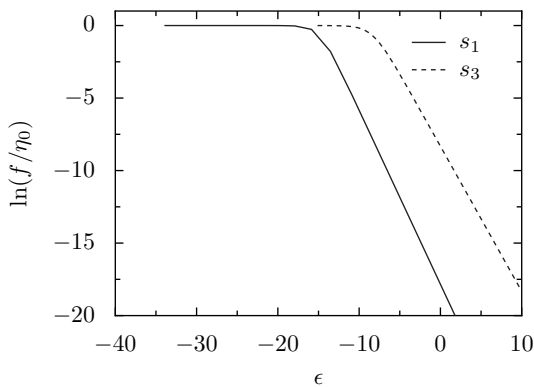,width=0.40\textwidth} } }
  \caption{The phase-space densities of $s_1$ and $s_3$ as functions
    of the energy per unit mass $\epsilon$. The functional form of
    these densities is the well-known Fermi-Dirac distribution, given by
    Eqs.~(\ref{eq:FD1}, \ref{eq:FD2}) and Table.~1}
  \label{fig:fs-LB}
\end{figure}

Figure~\ref{fig:rhos-LB} shows the radial density profiles of $s_1$
and $s_3$. As the $s_3$ configuration is more energetic, its mean
kinetic energy is higher and as a result the distribution is less
concentrated than the $s_1$ distribution, and has a lower density
core. Figure~\ref{fig:fs-LB} shows the DFs of $s_1$ and $s_3$. Notice
how both distributions have a substantial degenerate part, as for
both configurations the Fermi energies are $\epsilon_{\alpha_i} \simeq
\Phi_i(0)/2$.

\begin{figure}
  \centerline{ \hbox{
      \epsfig{file=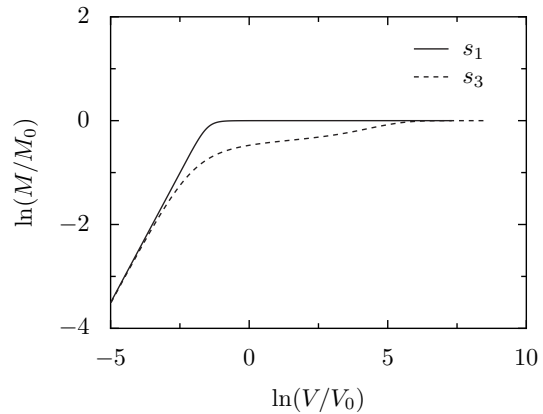, width=0.40\textwidth} } }
  \caption{The $M(V)$ functions of $s_1$ and $s_3$. $M_0=1$ is the
    total mass and $V_0=1$ is the initial phase-space volume that is
    occupied by the density level $\eta_0$. We notice that $M_3(V)\le
    M_1(V)$ for every $V$ and therefore $s_3$ is more mixed than $s_1$
    according to the mixing-theorem.}
  \label{fig:MVs-LB}
\end{figure}

Finally, to verify that the $s_3$ configuration is indeed more mixed
than the $s_1$ configuration - and therefore a transition $s_1\to s_3$
is permitted by the mixing theorem \citep{ref:Tre86} - we have
calculated the $M(V)$ functions of $s_1$ and $s_3$. The $M(V)$
function is defined in the following way: we first define $M(\eta)$
and $V(\eta)$ as the cumulative mass and phase-space volumes above the
phase-space density $\eta$,
\begin{eqnarray}
\label{def:VF}
  V(\eta) &\EqDef& 
      \int_{r<R}\!\! d^6\tau \, H[f(\Br,\Bv)-\eta] 
     = \int_\eta^\infty\!\! \tau(\eta')\, d\eta' \ ,\\
\label{def:MF}
  M(\eta) &\EqDef& \int_{r<R}\!\! d^6\tau\, H[f(\Br,\Bv)-\eta] 
     = \int_\eta^\infty\!\! \eta'\tau(\eta')\, d\eta' \ ,
\end{eqnarray}
with $H(\cdot)$ being the Heaviside step function.  As $V(\eta)$ is a
monotonically decreasing function, we can invert it and express $\eta$
in terms of the cumulative volume $V$.  Plugging this into $M(\eta)$
we get the $M(V)$ function. According to the mixing theorem, the
distribution $s_3$ is more mixed than the distribution $s_1$ if and
only if $M_3(V) \le M_1(V)$ for every $V$, as is shown in
\Fig{fig:MVs-LB}.

\begin{table*}
 \label{tab:s1-s3}
 \caption{The numerical parameters that specify the $s_1$ and $s_3$
   states. $E$ is the total energy of the state, $\eta_0$ is the
   initial density level given by \Eq{eq:eta}, $\beta, \mu$ are the
   Lagrange multipliers in Eqs.~(\ref{eq:FD1}, \ref{eq:FD2}), and
   $\Phi(0)$ is the gravitational potential at $r=0$.}
  \begin{tabular}{rrrrr}
  \hline
   State     & $E$  & $\eta_0\beta$ & $\mu$    & $\Phi(0)$ \\
   \hline\hline
   $s_1$     & $-6.157$ & $1.2$     & $-14.87$ & $-33.96$ \\
   $s_2$     & $-1.589$ & $1.0$     & $-8.322$ & $-15.12$ \\
   \hline
 \end{tabular}
\end{table*}

\subsection{Analysing the $s_2$ configuration}
\label{sec:LB-s2}

Let us now turn our attention to the $s_2$ configuration. Seemingly,
we need to calculate the $\tau_1(\eta)$ function of the $s_1$
configuration, and together with an energy of $E+\Delta E$, solve the
equations
\begin{eqnarray}
\label{eq:LB3-f}
  f_2(\epsilon) &=& \frac{1}{\Delta\eta}A(\epsilon)\int_0^{\eta_0}\!\! 
      \eta e^{-\beta_2 \eta(\epsilon-\mu_\eta)}\, d\eta \ , \\
\label{eq:LB3-A}
  A(\epsilon) &=& \frac{1}
     {1 + \frac{1}{\Delta\eta}\int_0^{\eta_0}\!\! 
        e^{-\beta_2 \eta(\epsilon-\mu_\eta)}\, d\eta} \ , \\
\label{eq:LB3-vol}
  \tau_1(\eta) &=& \frac{1}{\Delta\eta}\int_{\Phi_2(0)}^\infty\!\! 
      g_2(\epsilon) A(\epsilon) 
      e^{-\beta_2 \eta(\epsilon-\mu_\eta)}\, d\epsilon \ .
\end{eqnarray}
The function $g_2(\epsilon)$ needs to be calculated from \Eq{def:g}
using the gravitational potential $\Phi_2(r)$, which has to be
recovered from $f_2(\epsilon)$ using the Poisson
equation~(\ref{eq:LB2-possion}).  Finally, the resultant
$f_2(\epsilon)$ would be compared to $f_3(\epsilon)$ to see if the two
configurations are equal.

There is, however, a much simpler way to see if
$f_2(\epsilon)=f_3(\epsilon)$.  Let us assume that indeed this is the
case, and that consequently also $\Phi_2(r)=\Phi_3(r)$ and
$g_2(\epsilon)=g_3(\epsilon)$. In such case it is possible to recover
the full expression $\frac{1}{\Delta\eta}A(\epsilon)e^{-\beta_2
  \eta_0 (\epsilon-\mu_\eta)}$ in terms of the known functions
$f_3(\epsilon)$ and $g_3(\epsilon)$.

We start by replacing $f_2(\epsilon)\leftrightarrow f_3(\epsilon)$ and
$g_2(\epsilon)\leftrightarrow g_3(\epsilon)$ in
Eqs.~(\ref{eq:LB3-f}-\ref{eq:LB3-vol}). Differentiating \Eq{eq:LB3-A}
with respect to $\epsilon$ and substituting $f_3(\epsilon)$ from
\Eq{eq:LB3-f}, we obtain
\begin{equation}
  A'_2(\epsilon) = \beta_2 A(\epsilon) f_3(\epsilon) \ ,
\end{equation}
which yields
\begin{equation}
  A(\epsilon) = a_0 \exp\left(-\beta_2\int_\epsilon^\infty\!\!
    f_3(\epsilon')d\epsilon'\right) \EqDef a_0A_0(\epsilon) \ .
\end{equation}
Here $a_0$ is an unknown integration constant. The integral in
$A_0(\epsilon)$ can be easily done analytically if we recall the
definition of  $f_3(\epsilon)$ which is given in \Eq{eq:FD2},
yielding
\begin{equation}
  A_0(\epsilon) =
  \left[1+e^{-\beta_3\eta_0(\epsilon-\mu_3)}\right]^
      {-\frac{\beta_2}{\beta_3}} \ .
\end{equation}

Having found $A(\epsilon)$, we use \Eq{eq:LB3-vol} to find the
Lagrange multipliers $\mu_\eta$:
\begin{equation}
  e^{-\beta_2 \eta \mu_\eta} = \big[\tau_1(\eta)\big]^{-1} a_0
  \frac{1}{\Delta\eta}
  \int_{\Phi_3(0)}^\infty\!\! g_3(\epsilon) A_0(\epsilon) 
  e^{-\beta_2 \eta\epsilon}\, d\epsilon \ .
\end{equation}
Therefore 
\begin{equation}
  A(\epsilon)\frac{1}{\Delta\eta} 
     e^{-\beta_2\eta(\epsilon-\mu_\eta)} 
  = \frac{\tau_1(\eta)}{B_0(\eta)}A_0(\epsilon)
     e^{-\beta_2 \eta\epsilon} \ ,
\end{equation}
with 
\begin{equation}
\label{def:B}
 B_0(\eta) \EqDef \int_{\Phi_3(0)}^\infty\!\! 
           g_3(\epsilon') A_0(\epsilon') 
           e^{-\beta_2 \eta\epsilon'}\, d\epsilon' \ .
\end{equation}
This finally gives us
\begin{equation}
\label{eq:LB-final-f2}
  f_2(\epsilon) = A_0(\epsilon) \int_0^{\eta_0}\!\!
     \frac{\tau_1(\eta)}{B_0(\eta)}
     e^{-\beta_2 \eta\epsilon}\, d\eta \ .
\end{equation}
Note that the unknown integration constant $a_0$ has been cancelled
out. The only remaining unknown is $\beta_2$ which can be fixed by
requiring that the energy of $s_2$ will be equal to $E+\Delta E$.
Once this is done, we have an expression for $f_2(\epsilon)$ which is
equal to $f_3(\epsilon)$ if and only if $s_2$ is identical to $s_3$.

The procedure above is mathematically straightforward, however,
numerically it is slightly more complicated as the $\tau_1(\eta)$ has
a very strong peak near $\eta=\eta_0$ due to the degeneracy. It is
therefore preferable to perform the calculation using the cumulative
version of $\tau_1(\eta)$, which was defined in~(\ref{def:VF}). For a
spherical, isotropic system in a sphere of radius $R$ with a DF
$f(\epsilon)$ and a gravitational potential $\Phi(r)$, it is easy to
verify that $V(\eta)$ is given by
\begin{equation}
\label{eq:Vf-e}
  V(\eta) = \frac{2^{3/2}(4\pi)^2}{3} \int_0^{r(\eta)} \!\!
        s^2 [\epsilon(\eta)-\Phi(s)]^{3/2} \, ds \ ,
\end{equation}
with $\epsilon(\eta)$ being the inverse function of $f(\epsilon)$, and
$r(\eta)$ is
\begin{equation}
  r(\eta) \EqDef \left\{ \begin{array}{lcl}
         \Phi^{-1}[\epsilon(\eta)] &,& \epsilon(\eta) < -GM/R \\
         R                         &,& \epsilon(\eta) \ge -GM/R
       \end{array}\right. \ .
\end{equation}
Once $V_1(\eta)$ is calculated from the formula above [using
$f_1(\epsilon)$ and $\Phi_1(\epsilon)$], we can calculate
$f_2(\epsilon)$ from \Eq{eq:LB-final-f2} using integration by parts:
\begin{equation}
\label{eq:LB-by-parts-f2}
  f_2(\epsilon) = A_0(\epsilon)\int_0^{\eta_0}\!\! 
   V_1(\eta) \frac{e^{-\beta_2 \eta\epsilon}}{B_0(\eta)}
     \Big[1 -\beta_2 \eta\epsilon 
     - \eta\frac{B'_0(\eta)}{B_0(\eta)}\Big] \, d\eta  \ .
\end{equation}

\subsection{Results}
   
\begin{figure}
  \centerline{ \hbox{
      \epsfig{file=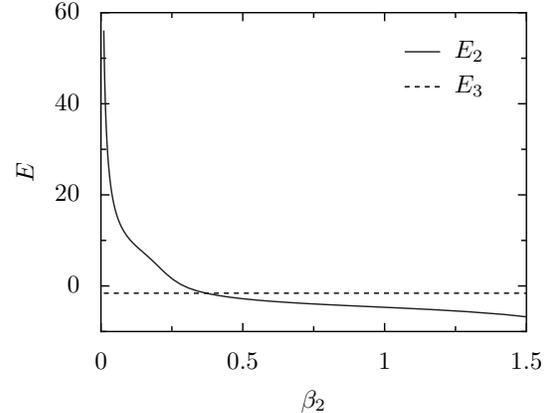,width=0.40\textwidth} } }
  \caption{The energy $E_2$ of different values of $\beta_2$, together
    with $E_3=-1.589$. $\beta_2=0.3707$ is the value that gives
  $E_2=E_3$. For higher values of up to $\beta_2=100$ we did not
  find any other solution and therefore we believe that the above
  solution is the only physical solution.}
  \label{fig:E2-LB}
\end{figure}

To satisfy the energy constraint $E_2 = E_3=-1.589$, we calculated
$E_2$ for various values of $\beta_2$ and chose $\beta_2=0.37071$
which gives the correct energy as shown in \Fig{fig:E2-LB}. We did not
find any other solution in the range $0.1<\beta<100$ and therefore we
believe that $\beta_2=0.37071$ is the only relevant solution.

\begin{figure}
  \centerline{ \hbox{
      \epsfig{file=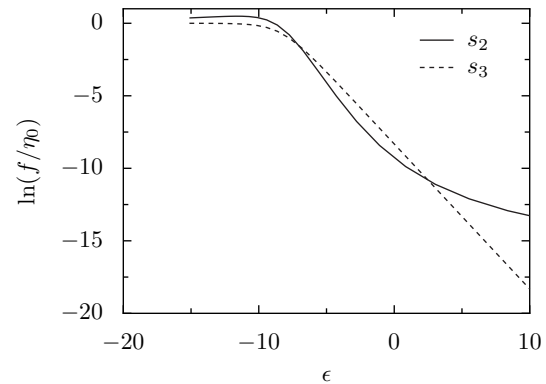,width=0.40\textwidth} } }
  \caption{Comparing $f_3(\epsilon)$ to $f_2(\epsilon)$ which was
    derived using \Eq{eq:LB-by-parts-f2}. The two functions are
    unequal proving that the $s_3$ state is different from the $s_2$
    state.  }
  \label{fig:f2-LB}
\end{figure}

Figure~\ref{fig:f2-LB} shows the graphs of $f_2(\epsilon)$ and
$f_3(\epsilon)$ once $\beta_2$ was fixed. Clearly the two graphs
strongly disagree, in some places by more than one order of magnitude
- much more than the numerical error in our calculations. The
conclusion is therefore that the LB67 theory is not transitive.

\newpage

\section{The Information-theory approach to violent relaxation and its
  relation to the LB67 theory}
\label{sec:info}

Recently, a new approach to violent relaxation was proposed in a
interesting paper by T.~K.~Nakamura \citep{ref:Nak00}. In that paper,
Nakamura uses an information-theory approach \citep{ref:Jay57a,
  ref:Jay57b} to define the entropy of a collisionless system and
thereby find its equilibrium state. Nakamura's theory (hereafter NK00)
predicts a different equilibrium state than LB67, and it is therefore
interesting to check weather his theory is transitive or not.  We will
not, however, try to answer the question which one of these theories
is more correct as it is, in our opinion, still an open question.
Instead, we shall first give a brief description of NK00 and its main
results, and then re-derive it theory using a combinatorial approach
which would enable us to compare it with the LB67 theory, and point to
the reasons of why they differ. Finally we will analyse the two-levels
configuration which goes into the water-bag configuration in a
limiting case. The result of this analysis will be used in the next
section when we examine the transitivity of the NK00 theory in a
double relaxation experiment.

\subsection{An outline of the NK00 theory}
\label{sec:NK00-outline}

In the NK00 theory we adopt the probabilistic description of the
phase-space density $f(t, \Br,\Bv)$. Let $f_0(\Br,\Bv)$ be the initial
phase-space density of the system, and define the initial probability
distribution $p_0(\Br,\Bv)$ for finding a (single) test point at $t=0$
by
\begin{equation}
  p_0(\Br,\Bv) = \frac{1}{M}f_0(\Br,\Bv) \ .
\end{equation}
Then we let the test point move under gravity just like any
phase-space element, and we define the probability distribution
$p(\Br,\Bv,t)$ as the probability distribution of finding the test
point at time $t>0$. The conservation of phase-space volume guarantees
that $p(\Br,\Bv,t)=f(\Br,\Bv,t)/M$ for all $t>0$.

Next, we divide phase-space into macro-cells $i=1,2,3,\ldots$ of
volume $\tw$, and define the coarse-grained probability $\bar{p}_i$ as
the probability of finding the point in the $i$'th macro-cell when
the system reaches an equilibrium. From the above discussion it is
clear that $\bar{p}_i$ is equal to $\bar{f}_i/M$ with $\bar{f}_i$
being the coarse-grained DF in the macro-cell $i$ at equilibrium.

To calculate $\bar{p}_i$ using the information-theory approach, we
define the \emph{joint probability-distribution} $p_i(\Br,\Bv)$ which
measures the probability of initially finding the test point at the
$(\Br,\Bv)$ and later at the macro-cell $i$. Then we maximise the
Shanon entropy
\begin{equation}
  S = -\sum_i\int\!\!d^6\tau p_i(\Br,\Bv)\log p_i(\Br,\Bv) \ ,
\end{equation}
subject to constraints of energy conservation, phase-space volume
conservation and initial conditions. The resultant distribution can be
best written in terms of the \emph{conditional probability}
$K_i(\Br,\Bv) = p_i(\Br,\Bv)/p_0(\Br,\Bv)$:
\begin{equation}
\label{eq:NK-K}
  K_i(\Br,\Bv) = e^{-\beta\epsilon_i - \delta(\Br,\Bv) 
      - \lambda_i/p_0(\Br,\Bv)} \ ,
\end{equation}
with $\epsilon_i$ being the energy per unit mass of the macro-cell
$i$, and $\beta, \delta(\Bx,\Bv), \lambda_i$ are the Lagrange
multipliers, to be found from the energy conservation constraint, from
the initial conditions
\begin{equation}
\label{eq:NK-con1}
  \sum_i K_i(\Br,\Bv) = 1 \ ,
\end{equation}
and from the phase-space volume conservation constraint
\begin{equation}
\label{eq:NK-con2}
  \int\!\! K_i(\Br,\Bv)\, d^6\tau = \tw \ .
\end{equation}
 
From Eqs.~(\ref{eq:NK-K}-\ref{eq:NK-con2}) it is evident that the
dependence of $K_i(\Br,\Bv)$ on the indices $i, \Br, \Bv$ is only via
$\epsilon_i$ and $p_0(\Br,\Bv)$, the latter can be trivially be
replaced by $f_0(\Br,\Bv)$. Therefore, the above equations can be
re-written using the $K(\epsilon, \eta)$ function, together with the
$\tau(\eta)$ and $g(\epsilon)$ functions which were defined in
Sec.~\ref{sec:LB-res}:
\begin{eqnarray}
\label{eq:NK2-K}
  K(\epsilon, \eta) &=& (\tw)^{-1}e^{-\beta\epsilon - \delta(\eta) -
                           \lambda(\epsilon)/\eta} \ , \\
\label{eq:NK2-con1}
  \int_{\Phi(0)}^\infty\!\! g(\epsilon) 
        K(\epsilon, \eta)\, d\epsilon &=& 1 \ , \\
\label{eq:NK2-con2}
  \int_0^\infty\!\! \tau(\eta) K(\epsilon, \eta)\, d\eta &=& 1 \ .
\end{eqnarray}
The coarse-grained equilibrium DF is then given by
\begin{equation}
\label{eq:NK2-f}
  f(\epsilon) = \int_0^\infty\!\! 
     \tau(\eta) \eta K(\epsilon, \eta)\, d\eta \ .
\end{equation}

As noted by Nakamura, a prominent difference between his result and
the LB67 results is that in the non-degenerate limit his expression
reduces to a single Maxwellian distribution, whereas the LB67
expression is a superposition of Maxwellian distributions with
different dispersions. This difference can be attributed to the fact
that in LB67 we discretise phase-space using phase-space elements of
equal volume and \emph{different masses}, while, as we shall see
below, the NK00 theory can be derived by discretising phase-space
using elements of equal mass, which are associated with different
phase-space volumes.

Another evident difference comes from the phase-space volume
conservation constraint \Eq{eq:NK-con1}. This constraint guarantees
that the total phase-space volume of all phase-space patches that
ended up in macro-cell $i$ will be equal to the macro-cell volume.
Consequently the total phase-space volume of the initial system
$\int_0^\infty\!\! \tau(\eta)\, d\eta$ must be equal to the total
phase-space volume of the non-vanishing phase-space density in the
equilibrium configuration. This constraint does not exist in the LB67
theory where the macro-cells can be only partly full - as is the case,
for example, in the non-degenerate equilibrium of a system which is
initially in the water-bag configuration. Furthermore, this can not be
trivially changed by adding a volume of zero phase-space density to
the initial condition, because setting $p(\Br,\Bv)=0$ would lead to
divergences in \Eq{eq:NK-K}. In Sec.~\ref{sec:NK-2-levels} we shall
see how this problem can be overcome by using a limiting procedure.

\subsection{Deriving the NK00 theory in a combinatorial approach}

To derive the NK00 theory in a combinatorial approach, we realise the
phase-space density distribution using $N\gg 1$ elements of
\emph{equal mass} $m$. As in Sec.~\ref{sec:LB-res}, we assume that
initially the system is made of a discrete set of density levels
$\eta_1, \eta_2, \ldots$ occupying phase-space volumes $V_1, V_2,
\ldots$. Then the overall number of elements that realise a
phase-space density $\eta_J$ is $N_J = V_J\eta_J/m$.

Next, we let the system reach an equilibrium through the process of
violent relaxation, and divide phase-space into macro-cells of equal
volume $\tw$, which are label by the index $i=1,2,\ldots$. We define a
\emph{micro-state} by specifying the macro-cell in which every
element ended up. A \emph{macro-state} is then defined by the matrix
$\{ n_{iJ}\}$ which counts how many elements that initially realised
the density level $\eta_J$ ended up in the macro-cell $i$. Using
$\{n_{iJ}\}$, the coarse-grained DF at macro-cell $i$ is given
by
\begin{equation}
\label{eq:NK-f}
  \bar{f}_i = \frac{m}{\tw}\sum_J n_{iJ} \ .
\end{equation}

Finally, we define the function $W\big(\{n_{iJ}\big)$ which counts how
many micro-states give the macro state $\{n_{iJ}\}$. It is then a
simple combinatorics to show that
\begin{equation}
  W\big(\{n_{iJ}\}\big) = \prod_J\frac{N_J!}{\prod_i n_{iJ}!} \ .
\end{equation}
Let us pause here and explain that this rather simple formula is a
result of the way we define a micro-state - by specifying the
macro-cell in which each element is found. We do not care exactly how
the different elements are distributed in each macro-cell. A different
approach, in which the macro-cells can be only partly full, and the
distribution of the different elements in a macro-cell is taken into
account when defining a micro-state, was taken by \citet{ref:Kul97}.
It is not difficult to see that when one adds the constraint that all
macro-cells must be completely full to their theory, Nakamura's
results are recovered. This is because in such case the number of
different ways to arrange the different elements in the macro-cell is
independent of which elements we are organising - as long as the
macro-cell is completely full. Therefore the number of micro-states in
a macro-state would be proportional to Nakamura's
$W\big(\{n_{iJ}\}\big)$, and consequently the equilibrium state would
be identical in both theories.

Next, we use $W\big(\{n_{iJ}\}\big)$ to define the entropy
\begin{equation}
  S \EqDef \log W \simeq const - \sum_{i,J}n_{iJ}(\log n_{iJ} - 1) \ ,
\end{equation}
where in the second equality we have used Stirling's formula to
approximate $\log(n_{iJ}!) \simeq n_{iJ}(\log n_{iJ} - 1)$.

Before maximising the entropy to find the most probable macro-state,
we first write down the constraints on $\{n_{iJ}\}$. The first
constraint comes from the initial conditions
\begin{equation}
\label{eq:NK3-con1}
  \sum_i n_{iJ} = N_J = \frac{V_J\eta_J}{m} \ .
\end{equation}
Then we have the phase-space volume conservation constraint, ensuring
that the total phase-space volume that is carried by elements that
ended up in the macro-cell $i$ will be exactly $\tw$, or, in other
words, that each macro-cell is completely filled:
\begin{equation}
\label{eq:NK3-con2}
  \sum_J n_{iJ} \frac{m}{\eta_J} = \tw \ .
\end{equation}
The last constraint is the energy constraint
\begin{equation}
\label{eq:energy-con}
  \sum_i \tw\bar{f}_i\left[\frac{1}{2}v_i^2 -
     \frac{1}{2}G\sum_j\frac{\tw\bar{f}_j}{|r_i - r_j|}\right] = E \ ,
\end{equation}
where $r_i$ and $v_i$ are the mean position and velocity of the $i$'th
macro-cell.

To maximise the entropy under the above constraint we use Lagrange
multipliers. The function that we wish to maximise with respect to
$n_{iJ}$ is therefore
\begin{eqnarray}
  I = S - \sum_J \lambda_i n_{iJ} - \sum_i \delta_J n_{iJ} 
           - \beta E \ .
\end{eqnarray}
Differentiating $I$ with respect to $n_{iJ}$ and equating it to zero,
we get
\begin{equation}
  \frac{dI}{dn_{iJ}} = -\log n_{iJ} - \delta_J -
  \frac{\lambda_i}{\eta_J} - \beta m \epsilon_i = 0 \ ,
\end{equation}
with $\epsilon_i = v_i^2/2 + \Phi(r_i)$ as usual, 
and therefore
\begin{equation}
\label{eq:NK3-n}
  n_{iJ} = e^{-\beta m \epsilon_i - \delta_J -
        \frac{\lambda_i}{\eta_J}} \ .
\end{equation}

Finally, we pass to a continuous description by giving every initial
phase-space density level a small width $\Delta \eta$. Then using the
$\tau(\eta)$ and $g(\epsilon)$ function we replace
\begin{eqnarray}
  \sum_J &\to& \frac{1}{\Delta \eta}\int\!\! d\eta \ , \\
  \sum_i &\to& \frac{1}{\tw}\int\!\! d\epsilon\,g(\epsilon) \ ,\\
  V_J &\to& \Delta \eta \tau(\eta_J) \ , \\
  n_{iJ} &\to& \frac{\Delta \eta\tw}{m}\,\tau(\eta_J)
         \eta_J K(\epsilon_i, \eta_J) \ .
\end{eqnarray}
Plugging these replacements into Eqs.~(\ref{eq:NK3-n},
\ref{eq:NK3-con1}, \ref{eq:NK3-con2}) and redefining 
$m\beta \to \beta$, we recover the NK00 Eqs.~(\ref{eq:NK2-K},
\ref{eq:NK2-con1}, \ref{eq:NK2-con2}).

This combinatorial formulation of the NK00 theory is very much along
the lines of ordinary statistical mechanic of a classical Boltzmann
gas.  Indeed, if we replace the notion of phase-elements with
particles of equal mass and discard the constraint of conservation of
phase-space volume \Eq{eq:NK2-con2}, we have a text-book derivation of
the Boltzmann gas statistics. It is therefore not surprising that
Nakamura found that his equilibrium DF reduces to the well-known
Maxwell-Boltzmann distribution in such case. We do not agree, however,
with Nakamura's claim that this property is a proof for its
correctness over the LB67 theory. This is because a collisionless
relaxation is essentially a very different process from the
collision-full relaxation that occurs in Boltzmann gas, driven by
different physical processes over different timescales. However, as
previously mentioned, deciding which theory is more correct is not the
goal of this paper.

\subsection{Analysing the two-levels configuration}
\label{sec:NK-2-levels}

As was noted in the end of Sec.~\ref{sec:NK00-outline}, the NK00
theory cannot handle a zero phase-space density directly. Therefore it
is not straightforward to analyse the equilibrium state that results
from an initial water-bag configuration, as in this configuration
there is one patch of phase-space density $\eta_0$ surrounded by an
infinite volume of zero phase-space density. The way this can be done
is to consider an initial state with two density levels $\eta_0$ and
$\eta_1$ with corresponding volumes $V_0$ and $V_1$. The water-bag
configuration is then recovered by taking the limit $\eta_1\to 0,
V_1\to\infty$ and $\eta_1 V_1 \to 0$.

To derive the equilibrium configuration of the two-levels system we
use the fact that in this particular case the matrix $\{n_{iJ}\}$ can
be expressed in terms of $\bar{f}_i$, thereby greatly simplifying the
end result. Let us then re-derive the equilibrium equation for this
particular case instead of using Eqs.~(\ref{eq:NK2-K}-\ref{eq:NK2-f}).
Denoting by $n_{i,0}$ and $n_{i,1}$ the total number of elements of
$\eta_0$ and $\eta_1$ that end-up in the $i$'th macro-cell, the
coarse-grained DF is given by
\begin{equation}
\label{eq:2lvl-f}
  \bar{f}_i = \frac{m}{\tw}(n_{i,0} + n_{i,1}) \ .
\end{equation}
Then using the conservation of phase-space volume constraint~(\ref{eq:NK3-con2}),
\begin{equation}
\label{eq:2lvl-vol}
   n_{i,0}\frac{V_0}{N_0} + n_{i,1}\frac{V_1}{N_1} = \tw \ ,
\end{equation}
with $N_0, N_1$ being the total number of elements with densities
$\eta_0, \eta_1$, given by \Eq{eq:NK3-con1}, we express $n_{i,0}$ and
$n_{i,1}$ in terms of $\bar{f}_i$:
\begin{eqnarray}
  n_{i,0} &=& \frac{\tw}{m}\frac{\bar{f}_i-\eta_1}{\eta_0-\eta_1}\eta_0 \ , \\
  n_{i,1} &=& \frac{\tw}{m}\frac{\eta_0-\bar{f}_i}{\eta_0-\eta_1}\eta_1 \ .
\end{eqnarray}
The energy constraint is given by \Eq{eq:energy-con}, and the
initial-condition constraint is
\begin{equation}
\label{eq:2lvl-N2}
  \sum_i n_{i,0} = N_0 = \frac{\eta_0 V_0}{m} \ .
\end{equation}
Notice that we need only the $N_0$ constraint since the $N_1$
constraints follows directly from requiring that the total phase-space
volume occupied by the equilibrium system would be equal to $V=V_0 +
V_1$. In fact, instead of \Eq{eq:2lvl-N2}, we can use an alternative
\emph{total mass} constraint, provided that the overall phase-space
volume is conserved. This is done as follows: expressing $n_{i,0}$ in
terms of $\bar{f}_i$ in \Eq{eq:2lvl-N2} we get 
\begin{equation}
  \tw\sum_{i}\frac{\bar{f}_i-\eta_1}{\eta_0-\eta_1} = V_0 \ ,
\end{equation}
which gives us
\begin{equation}
  \sum_i \tw \bar{f}_i - \eta_1\sum_i \tw = V_0 \eta_0 - V_1 \eta_1 \ .
\end{equation}
But $\sum_i \tw = V_0 + V_1$ (conservation of total phase-space
volume) and therefore we find
\begin{equation}
  \sum_i \tw \bar{f}_i = V_0 \eta_0 + V_1 \eta_1 = M \ .
\end{equation}

Adding these constraints together with the appropriate Lagrange
multipliers to the entropy, the expression that we need to maximise is
\begin{eqnarray}
  I &\simeq& const - \sum_{i}n_{i,0}\big(\log n_{i,0} - 1\big) \\
    && - \sum_{i}n_{i,1}\big(\log n_{i,1} - 1\big)
    + \tilde{\beta}E
    + \tilde{\mu}\sum_i\tw \bar{f}_i \ .
\end{eqnarray}
Differentiating $I$ with respect to $\bar{f}_i$ and equating to $0$
we find
\begin{eqnarray}
  &&-\frac{\eta_0}{\eta_0-\eta_1}\frac{\tw}{m}
     \log\left(\frac{\tw}{m}\frac{\bar{f}_i-\eta_1}{\eta_0-\eta_1}\eta_0\right) \\
  &&+\frac{\eta_1}{\eta_0-\eta_1}\frac{\tw}{m} 
     \log\left(\frac{\tw}{m}\frac{\eta_0-\bar{f}_i}{\eta_0-\eta_1}\eta_1\right)
  + \tilde{\beta}\tw\epsilon_i + \tilde{\mu}\tw = 0 \ .
\end{eqnarray}
After a trivial algebra and redefinition of the Lagrange multipliers
$\tilde{\beta}$ and $\tilde{\mu}$ to $\beta$ and $\mu$, 
we obtain
\begin{equation}
\label{eq:2lvl-f-final}
  \frac{\frac{\bar{f}_i-\eta_1}{\eta_0-\eta_1}}
     {\left(\frac{\eta_0-\bar{f}_i}{\eta_0-\eta_1}\right)^{\eta_1/\eta_0}}
  = e^{-\beta(\epsilon_i-\mu)} \ .
\end{equation}
Notice how the denominator provides an upper cut-off for $\bar{f}_i$, as it
forbids it from exceeding $\eta_0$.

Consider now the $\eta_1\to 0$ limit. Seemingly, it would go into an
isothermal sphere
\begin{equation}
  \bar{f}_i = \eta_0 e^{-\beta(\epsilon_i - \mu)} \ ,
\end{equation}
but this is not the case as for every finite $\eta_1$, $\bar{f}_i$
cannot exceed $\eta_0$. It is easy to see that the right limit is
therefore
\begin{equation}
\label{eq:NK-FD}
  f_{NK}(\epsilon) = \left\{\begin{array}{lcl}
      \eta_0                      &,& \epsilon<\mu \\
      \eta_0e^{-\beta(\epsilon-\mu)} &,& \epsilon\ge \mu 
      \end{array}\right. \ .
\end{equation}
This distribution is not the LB67 Fermi-Dirac distribution given by
\Eq{eq:FD1} or \Eq{eq:FD2}, but is what corresponds to that
distribution on the NK00 theory. It is not smooth, and is exactly
isothermal for energies $\epsilon>\mu$.

For the water-bag model in LB67 the condition that no two elements of
phase-density can overlap leads to a statistics with exclusion,
equivalent to the Fermi-Dirac problem. It is not clear to us how
Nakamura's formulation could obtain the Fermi-Dirac statistics.

\section{Non-transitivity in the Nakamura theory of violent relaxation}
\label{sec:NK00}

Having found the equilibrium configuration of the water-bag initial
configuration in the NK00 theory, we are in a position to test the
theory's transitivity. The procedure for that is identical to the one
that was used in the LB67 case, in
sections~\ref{sec:double-relax},~\ref{sec:LB-s2}, and therefore will
not be repeated. Instead, we shall first describe how the $s_1$ and
$s_3$ configurations are found and then how $s_2$ is compared to the
$s_3$ configuration.

To find the $s_1$ and $s_3$ configurations, we must first find the
gravitational potential of the DF in \Eq{eq:NK-FD} in a sphere of
radius $R$, and then fix $\beta$ and $\mu$ so that the overall energy
and mass will be equal to $E$ and $M$. Additionally, just as in the
LB67 case, we assume that the final equilibrium state is spherical and
therefore the Poisson equation for $\Phi(r)$ is:
\begin{equation}
\label{eq:NK-possion}
  \frac{1}{r^2}\frac{d}{dr}
     \left(r^2\frac{d\Phi}{dr}\right) = (4\pi)^2 G \int_0^\infty v^2 
       f_{NK}\big[v^2/2 + \Phi(r)\big] \, dv\ .
\end{equation}
Passing from $\Phi(r)$ to the dimensionless $\psi(r)$ by
\begin{equation}
  \psi(r) \EqDef \beta[\mu - \Phi(r)] \ ,
\end{equation}
\Eq{eq:NK-possion} simplifies to
\begin{eqnarray}
  && -\frac{1}{\beta r^2}\frac{d}{dr}
     \left(r^2\frac{d\psi}{dr}\right)= (4\pi)^2 G \eta_0 \\
   &&\times \int_0^\infty\!\! v^2\,dv
    \left\{ \begin{array}{lcl} 
       1 &,& \beta v^2/2 < \psi(r) \\
       e^{\psi(r)} e^{-\beta v^2/2} &,&
           \beta v^2/2 \ge \psi(r) 
            \end{array} \right.
\end{eqnarray}
Finally, changing variables $r\to x$
\begin{equation}
    x \EqDef \left(\frac{16\pi^2
         \sqrt{2}G\eta_0}{\beta^{1/2}}\right)^{1/2}r \ , 
\end{equation}
and using the Error-function
$\mbox{erf}(x)=\frac{2}{\sqrt{\pi}}\int_0^x e^{-t^2}dt$, the ordinary
differential equation for $\psi(x)$ is
\begin{equation}
  -\frac{1}{x^2}\frac{d}{dx}
      \left(x^2\frac{d\psi}{dx}\right) = I(\psi) \ ,
\end{equation}
with
\begin{eqnarray}
  I(\psi) = 
  \left\{
      \begin{array}{lcl}
      \frac{\sqrt{\pi}}{2}e^{\psi} &,& \psi < 0 \\
      \frac{2}{3}\psi^{3/2} + \sqrt{\psi} 
       + \frac{\sqrt{\pi}}{2}e^{\psi}
         \left(1-\mbox{erf}\sqrt{\psi}\right) &,& \psi \ge 0
      \end{array}\right. \ .
\end{eqnarray}
To integrate this equation we must first set its initial condition. We
let $\psi(0)=\psi_0$ be a free parameter, and $\psi'(0)=0$ since in a
spherical system the gravitational force vanishes in the centre. Then
once $\psi(x)$ is (numerically) found, we fix $\mu$ by requiring that
$\Phi(R) = -GM/R$. This way we can find the gravitational potential,
and thereafter the total mass and energy for any given $\beta$ and
$\psi_0$. The last step is to find the right $\beta$ and $\psi_0$ that
would give us $M$ and $E$.

Practically, instead of looking for $\beta$, $\psi_0$ for a specific
$E$ and $M$ choice in $s_1$ and $s_3$, we have picked two different
values of $\beta$ and and two corresponding values of $\psi_0$ to
satisfy the total mass constraint. Once $\psi_0$ is fixed it also
defines a total energy. We chose $\beta_1>\beta_3$ in order to obtain
$E_1<E_2$.

Let us now see how the $s_2 \leftrightarrow s_3$ comparison can be
done. According to Sec.~\ref{sec:info}, $f_2(\epsilon)$ is determined
by the following set of equations:
\begin{eqnarray}
\label{eq:NK4-K}
  K(\epsilon, \eta) &=& (\tw)^{-1}e^{-\beta_2\epsilon - \delta(\eta) 
       - \lambda(\epsilon)/\eta} \ , \\
\label{eq:NK4-con1}
  1 &=& \int_{\Phi_2(0)}^\infty g_1(\epsilon) 
         K(\epsilon, \eta)\, d\epsilon \ , \\
\label{eq:NK4-con2}
  1 &=& \int_0^{\eta_0} \tau_1(\eta) 
         K(\epsilon, \eta)\, d\eta \ , \\
\label{eq:NK4-f2}
  f_2(\epsilon) &=& \int_0^{\eta_0} \tau_1(\eta) \eta
         K(\epsilon, \eta)\, d\eta \ .
\end{eqnarray}
Additionally, we know that $f_3(\epsilon)$ is given by
\begin{equation}
\label{eq:NK-f3}
  f_3(\epsilon) = \left\{\begin{array}{lcl}
      \eta_0                      &,& \epsilon<\mu_3 \\
      \eta_0e^{-\beta_3(\epsilon-\mu_3)} &,& \epsilon\ge \mu_3
      \end{array}\right. \ ,
\end{equation}
and $g_3(\epsilon)$, $\Phi_3(r)$ have been found as described above.
Assuming that $s_2=s_3$, we replace these functions with
$f_2(\epsilon), g_2(\epsilon)$ and $\Phi_2(\epsilon)$ in
Eqs.~(\ref{eq:NK4-con1}-\ref{eq:NK4-f2}), and differentiate
\Eq{eq:NK4-f2} with respect to $\epsilon$. Using \Eq{eq:NK4-con2} we
get
\begin{equation}
  f'_3(\epsilon) = -\beta_2 f_3(\epsilon) - \lambda'(\epsilon) \ ,
\end{equation}
and therefore
\begin{equation}
  \lambda(\epsilon) = C + \beta_2\int_\epsilon^\infty
      f_3(\epsilon')d\epsilon' - f_3(\epsilon) 
  \EqDef C + \lambda_0(\epsilon) \ ,
\end{equation}
with $C$ some unknown integration constant. The integral in
$\lambda_0(\epsilon)$ can be done analytically, yielding
\begin{equation}
  \lambda_0(\epsilon) = \left\{
    \begin{array}{lcl}
      \eta_0\left[ \frac{\beta_2}{\beta_3}-1 
          + \beta_2(\mu_3-\epsilon)\right] &,& \epsilon<\mu_3 \\
      \eta_0\left(\frac{\beta_2}{\beta_3}-1\right) 
        e^{-\beta_3(\epsilon-\mu_3)} &,& \epsilon \ge \mu_3 
    \end{array}\right. \ .
\end{equation}
Then from
\Eq{eq:NK4-con1} we find that
\begin{eqnarray}
\label{eq:Ks2}
  e^{\delta(\eta)} &=& e^{-C/\eta}(\tw)^{-1}\int_{\Phi_3(0)}^\infty\!\! g_3(\epsilon) 
   e^{-\beta_2\epsilon - \lambda_0(\epsilon)/\eta} \, d\epsilon  \\
   &\EqDef& e^{-C/\eta}(\tw)^{-1}D_0(\eta) \ ,
\nonumber
\end{eqnarray}
and therefore
\begin{eqnarray}
  K(\epsilon,\eta) &=& 
  (\tw)^{-1}e^{-\beta_2\epsilon 
         - \delta(\eta)-\lambda(\epsilon)/\eta} \\
  &=& \frac{e^{-\beta_2\epsilon 
         -\lambda_0(\epsilon)/\eta}}{D_0(\eta)} \ .
\end{eqnarray}
Notice how the unknown integration constant $C$ and the dimensional
constant $\tw$ are cancelled out.

Next we calculate $f_2(\epsilon)$ using \Eq{eq:NK4-f2} and fix
$\beta_2$ such that the energy of the system is $E+\Delta E$. Once
$\beta_2$ is fixed, $K(\epsilon, \eta)$ is completely resolved in
terms of $s_1$ and $s_3$ functions, and we can check if it solves the
maximum-entropy equations by plugging it into \Eq{eq:NK4-con2}.

Finally, we should note that as in the LB67 case, the $\tau_1(\eta)$
function has a strong peak at $\eta=\eta_0$ due to the degeneracy.
Here, however, this peak is proportional to $\delta(\eta-\eta_0)$ as
$f_1(\epsilon)=\eta_0$ for every $\Phi_1(0)<\epsilon<\mu_1$. The
prefactor in front of this delta function is $V_{deg}$ - the volume of
phase-space for which $\Phi_1(0)<\epsilon<\mu_1$, which can be easily
calculated from \Eq{eq:Vf-e}. Therefore, to preform the integration
over $\tau_1(\eta)$ in Eqs.~(\ref{eq:NK4-con2}, \ref{eq:NK4-f2})
numerically, we first calculate the smooth contribution which comes
from the $\tau_1(\eta)$ with $\eta<\eta_0$, and then add the
delta-function contribution by evaluating the integrands at
$\eta=\eta_0$ and multiplying them by $V_{deg}$.

\subsection{Numerical results}
\label{sec:NK-numres}

As in the LB67 case, the $s_0$ state was constructed as a water-bag
configuration with $\eta$ given by \Eq{eq:eta} and $M=1$, $R=1$,
$G=1$. The $s_1$ and $s_3$ configurations were then chosen as
described in the previous sub-section, by fixing $\beta$ and varying
$\psi_0$ until the total mass constraint was satisfied. The main
numerical parameters of these configurations are summarised in
Table~2.
\begin{table*}
 \label{tab:NK:s1-s3}
 \caption{The numerical parameters that specify the $s_1$ and $s_3$
   states in NK00 double relaxation experiment. $E$ is the total
   energy, $\beta$ and $\mu$ are the Lagrange multipliers in
   \Eq{eq:NK-FD}, and $\Phi(0)$ is the gravitational potential at $r=0$.
 }
  \begin{tabular}{rrrrr}
  \hline
   State     & $E$   & $\beta$ & $\mu$ & $\Phi(0)$ \\
   \hline\hline
   $s_1$     & $-6.405$ & $3.0$     & $-15.30$     & $-35.08$ \\
   $s_3$     & $-4.306$ & $0.68$    & $-14.08$     & $-27.80$ \\
   \hline
 \end{tabular}
\end{table*}

\begin{figure}
  \centerline{ \hbox{
      \epsfig{file=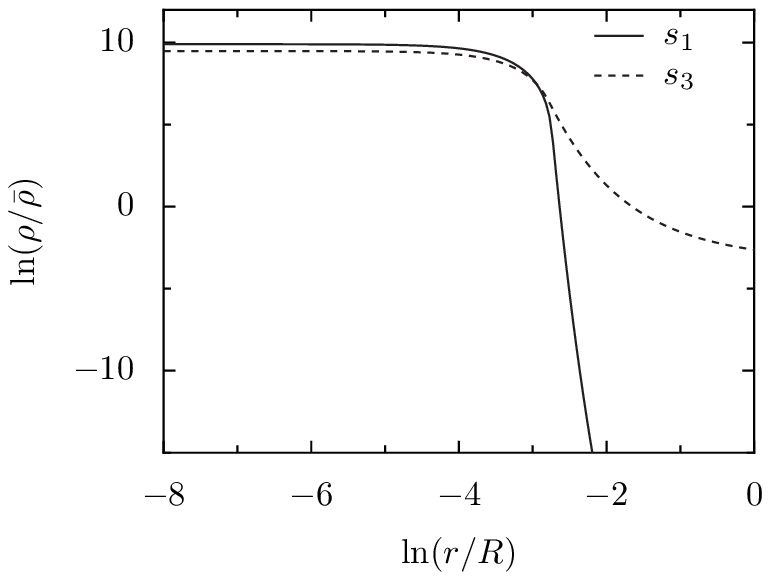,width=0.40\textwidth} } }
  \caption{Density profiles of the $s_1$ and $s_3$ states in the
    double-relaxation experiment of the NK00 theory. $\bar{\rho}$ is
    the average density given by $\bar{\rho}=3M/(4\pi R^3)$. As in
    \Fig{fig:rhos-LB}, the $s_3$ state corresponds to a hotter system
    with higher energy which makes it less concentrated.}
  \label{fig:rhos-NK}
\end{figure}

\begin{figure}
  \centerline{ \hbox{
      \epsfig{file=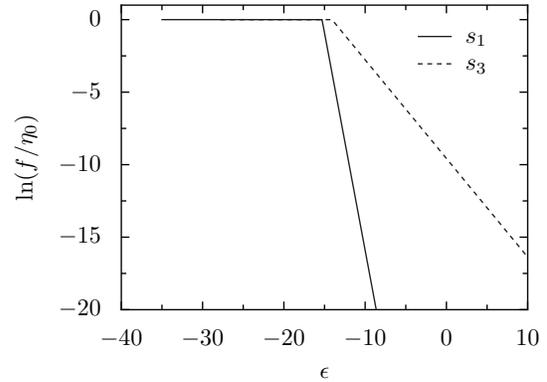,width=0.40\textwidth} } }
  \caption{DFs of $s_1$ and $s_3$ in the NK00 experiment. The
    functional form of these DFs is given in \Eq{eq:NK-FD} and
    Table.~2. Unlike the Fermi-Dirac DFs of the LB67
    theory given in \Fig{fig:fs-LB}, these DFs have a sharp transition
    between a completely degenerate core with $f=\eta_0$ for
    $\epsilon<\mu$ to an isothermal envelope for $\epsilon\ge\mu$.}
  \label{fig:fs-NK}
\end{figure}

\begin{figure}
  \centerline{ \hbox{
      \epsfig{file=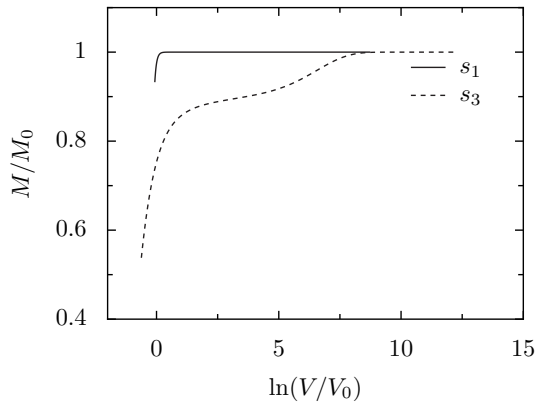,width=0.40\textwidth} } }
  \caption{The $M(V)$ functions of $s_1$ and $s_3$ in the
    double-relaxation experiment of the NK00 theory. $M_0$ and $V_0$
    are defined as in \Fig{fig:MVs-LB}, but here the $y$-axis is
    $M/M_0$ whereas in \Fig{fig:MVs-NK} is $\ln(M/M_0)$. This is
    because the range over which $M/M_0$ varies is much smaller than
    that of the NK00 case. Nevertheless, as in \Fig{fig:MVs-LB},
    $M_3(V)\le M_1(V)$ for every $V$, showing that $s_3$ is more mixed
    than $s_1$.}
  \label{fig:MVs-NK}
\end{figure}

Figure~\ref{fig:rhos-NK} shows the density profiles of the $s_1$ and
$s_3$ configuration. As expected, the hotter system, $s_3$, has a core
with a lower density than the $s_1$ system. Figure~\ref{fig:fs-NK}
shows the DF of these systems, and \Fig{fig:MVs-NK} compares the
$M(V)$ functions of these two states, showing that $M_3(V) \le M_1(V)$
for every $V$ - and therefore the transition $s_1 \to s_3$ is allowed.

\begin{figure}
  \centerline{ \hbox{
      \epsfig{file=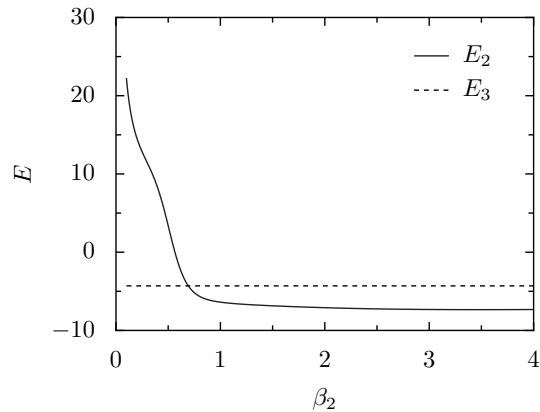,width=0.40\textwidth} } }
  \caption{The energy $E_2$ of $s_2$ as a function of $\beta_2$,
    together with the energy $E_3=-4.306$. The value of $\beta_2$ for
    which the two energies agree is $\beta_2=0.6944$. No other
    solutions were founds for $\beta_2$ up to $100$, and we therefore
    believe that this is the only physical solution.}
  \label{fig:E2-NK}
\end{figure}

Finally, to compare the $s_2$ configuration to the $s_3$ configuration
we have varied $\beta_2$ in the range $0.1<\beta_2<100$ until we
obtained $E_2=E_3=-4.306$ with $\beta_2=0.6944$. This was the only
solution in that range, and we believe that it is the only physical
solution in general. The graph of $E_2(\beta)$ verses $E_3$ in the
range $[0.1,4]$ is shown in \Fig{fig:E2-NK}.

\begin{figure}
  \centerline{ \hbox{
      \epsfig{file=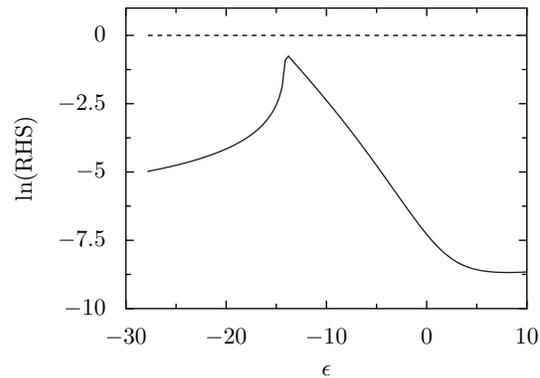,width=0.40\textwidth} } }
  \caption{Comparing the RHS of \Eq{eq:NK2-con2} to its LHS (in
    natural logarithmic scale). The two sides of the equations are in
    gross disagreement, proving that also the NK00 theory is
    non-transitive.}
  \label{fig:check-NK}
\end{figure}

Once $\beta_2$ was found, we used the expression of $K(\epsilon, \mu)$
in \Eq{eq:Ks2}, to calculate the RHS of \Eq{eq:NK4-con2} and compare
it to $1$. Figure~\ref{fig:check-NK} shows this comparison. The
disagreement between the RHS and $1$ is sometimes as high as 3 orders of
magnitude - much higher than any possible numerical error. We
therefore conclude that also the NK00 theory is non-transitive.

\section{Conclusions}
\label{sec:conc}

In this paper we have demonstrated that the statistic-mechanical
theories of violent relaxation by Lynden-Bell and Nakamura are both
non-transitive. This non-transitivity is a result of the phase-mixing
that occurs when the system relaxes; as the fine-grained phase-space
density filaments become thiner and thiner, the system is better
described in terms of the coarse-grained phase-space density.  Any
further relaxation of the system should be therefore considered in
terms of the coarse-grained phase-space density - which as we have
seen would yield different results from a prediction that is based on
the initial fine-grained phase-space density. This is a worrying
aspect of these theories as it is easy to imagine a scenario where
part of the system mixes, then fluctuates, and then mixes once again.
The predictions of the theory, based on the fine-grained density, will
then give us a wrong result.

In some sense we have been breaking into an open door. Even without
considering the non-transitivity of the theories, they are plagued
by severe problems. There exist two equally plausible ways of
discretising phase-space, one with equal volume elements and one with
equal mass elements, which yield two different results. More
importantly, the ability of the theories to predict the final
outcome of a violent-relaxation process is very limited. Indeed, as
was mentioned in \Sec{sec:overview}, the most important reason for
this is that violent-relaxation is almost never complete; the
fluctuations of the gravitational potential die much faster for the
system to settle in the most probable state.

Nevertheless, we believe that these difficulties and ambiguities in
exactly how to do the statistical mechanics of the collisionless
Boltzmann equation teach us an important lesson. The non-transitivity
that we have shown is a sign that a kinetic description of violent
relaxation is probably incomplete, as the equilibrium is dependent on
the evolutionary path of the system. Instead, what is probably needed
is a dynamical approach to the problem. Indeed most of the above
difficulties are circumvented if instead of aiming to derive a
universal most probable state, we reduce our aim to that of finding an
appropriate and useful evolution equation for the coarse-grained
$\bar{f}$. 

An interesting attempt to find such equation was taken by
\citet{ref:Cha98b}, who used the maximal entropy-production principle
(MEPP) to obtain a close equation for $\bar{f}$. His analysis,
however, uses the initial fine-grained $\tau(\eta)$ to define the
(Lynden-Bell) entropy rather than the instantaneous, coarse-grained
$\tau(\bar{\eta})$, which according to the above discussion is more
correct. Derivation of a useful dynamical equation for $\bar{f}$ thus
remains a challenging open problem.

\section*{Acknowledgments}
We thank Peter Johansson for his help with the manuscript.  This work
was supported by a Marie-Curie Individual Fellowship of the European
Community No. HPMF-CT-2002-01997.

\label{lastpage}

\end{document}